\documentclass{article}
\usepackage{amsmath, amssymb,graphicx, setspace, epstopdf}
\usepackage{cite}
\usepackage[margin=2.0cm]{geometry}

\begin{document}

\title{A squeeze-like operator approach to position-dependent mass in quantum mechanics}
\author{H{\' e}ctor M. Moya-Cessa$^{1}$, Francisco Soto-Eguibar$^{1}$, Demetrios N. Christodoulides$^{2}$\\
\small{$^{1}$Instituto Nacional de Astrof\'{i}sica, \'{O}ptica y Electr\'{o}nica INAOE} \\
\small{Luis Enrique Erro 1, Santa Mar\'{i}a Tonantzintla, San Andr\'{e}s Cholula, Puebla, 72840 M{e}xico}\\
\small{$^{2}$The College of Optics and Photonics, University of Central Floria CREOL, Orlando, Florida, USA}}
\date{}
\maketitle

\begin{abstract}
We provide a squeeze-like transformation that allows one to remove a position dependent mass from the Hamiltonian. Methods to solve the Schr\"{o}dinger equation may then be applied to find the respective eigenvalues and eigenfunctions. As an example, we consider a position-dependent-mass that leads to the integrable Morse potential and therefore to well-known solutions.
\end{abstract}

\section{Introduction}
Considerable interest has been recently devoted in finding exact solutions to Schr\"odinger equations involving known potentials when the mass is position-dependent (PDM). Among them, one may mention the Morse and Coulomb potentials \cite{1,2,3,4,5,6,7,8,9,10,11,12,13,14,15,16,17,18}. 
Moreover, it has been recently shown \cite{19} that to lowest order of perturbation theory, there exists a whole class of Hermitian position-dependent-mass Hamiltonians that are associated with pseudo-Hermitian Hamiltonians.

A great deal of interest has been paid to the interplay between these pseudo-Hermitian PT-symmetric Hamiltonians and their equivalent Hermitian representations \cite{20,21,22,23,24}. In particular, Mostafazadeh \cite{20,21} has considered the transition to the classical limit by showing that the relevant classical Hamiltonian for the PT-symmetric cubic anharmonic oscillator plus a harmonic term, produces a behavior similar to a point particle with position-dependent-mass interacting with a quartic harmonic oscillator.

Indeed, many physical settings exist in which the effective mass can in principle depend on position. For example, Wang et al. \cite{25} have recently shown that the Schr\"{o}dinger equation for a thin charged shell moving under the influence of its own gravitational field may be viewed as a position-dependent-mass problem.

Displacement operators have already been introduced for systems with position-dependent-mass, for null or constant potentials from which generalized forms of the momentum operator have been obtained \cite{26,27}.

In this contribution, we demonstrate the possibility of transforming via similarity transformations, a position dependent mass Hamiltonian into a Hamiltonian with constant (unity) mass. By doing so, these Hamiltonians can then be solved (if integrable) using well-known techniques from quantum mechanics. If on the other hand the potentials are not solvable, perturbative methods may be applied for their solution. In order to achieve this objective, we use aspects associated with some non-classical states of the harmonic oscillator, namely, squeezed states \cite{28,29}. For squeezed states, the uncertainty may be "squeezed" in one of the quadratures, while in the other canonical conjugate variable the uncertainty increases.

\section{Squeeze operator}
In what follows, we will first show how the constant mass may be eliminated from the kinetic energy in a Hamiltonian. In this regard, consider the Hamiltonian
\begin{equation}\label{1}
 \hat{H}=\frac{\hat{p}^2}{2 m_0}+V(x),
\end{equation}
where the mass particle is $m_0$ and $\hbar=1$. This Hamiltonian is in turn transformed using the squeeze unitary operator \cite{28}
\begin{equation}\label{2}
  \hat{R}=\exp\left[-i\frac{\ln m_0}{4}(\hat{p}\hat{x}+\hat{x}\hat{p})\right].
\end{equation}
To find how the operator $\hat{R}$  transforms the position and the momentum operators, the Hadamard lemma \cite{30} is used; i.e., that  $e^{\hat{A}} \hat{B} e^{-\hat{A}}=\hat{B}+\left[ \hat{A},\hat{B}\right] + \frac{1}{2!}\left[ \hat{A},\left[ \hat{A},\hat{B}\right] \right]+ \frac{1}{3!}\left[ \hat{A},\left[ \hat{A},\left[ \hat{A},\hat{B}\right]\right] \right]+...$, from which we obtain that
\begin{equation}\label{3}
	 	\hat{R} \hat{x} \hat{R}^\dagger=\frac{\hat{x}}{\sqrt{m_0}},\qquad  \hat{R} \hat{p} \hat{R}^\dagger={\sqrt{m_0}} \hat{p}.
\end{equation}
As a result, the transformed Hamiltonian takes the form
\begin{equation}\label{4}
  \hat{H}_\mathrm{R}=\hat{R} \hat{H} \hat{R}^\dagger=\frac{\hat{p}^2}{2}+V\left( \frac{x}{\sqrt{m_0}}\right),	 	
\end{equation}
and thus the mass has been effectively eliminated from the kinetic energy term. Based on this latter possibility, one could ask if the mass can also be eliminated from the kinetic energy via a proper transformation, even if it is position dependent.

\section{Position dependent mass}
There is always some uncertainty as to the actual form of the kinetic energy term in a Hamiltonian, when the mass is position dependent. This is because $m(x)$ no longer commutes with the momentum.  There are consequently several ways to write the kinetic part of the Hamiltonian that must be kept Hermitian; for instance
\begin{equation}\label{5}
   \hat{H}_\mathrm{kin}=\frac{1}{4} \left( m^\alpha \hat{p} m^\beta \hat{p} m^\gamma + m^\gamma \hat{p} m^\beta \hat{p} m^\alpha  \right), \qquad \alpha+\beta+\gamma=-1 .
\end{equation}
On the other hand, by choosing $\alpha=\gamma=0, \quad  \beta=-1$, we arrive to the ordering proposed by BenDaniel and Duke \cite{31},
\begin{equation}\label{6}
    \hat{H}_\mathrm{kin}=\hat{p}\frac{1}{2m\left(x\right) }\hat{p},
\end{equation}
while with the choice $\alpha=-1$, $\beta=\gamma=0$, we get
\begin{equation}\label{7}
     \hat{H}_\mathrm{kin}=\frac{1}{4}\left[ \frac{1}{m\left(x\right) }\hat{p}^2 + \hat{p}^2 \frac{1}{m\left(x\right) } \right].
\end{equation}
Although there is no apparent reason in selecting any particular ordering for the kinetic position-dependent-mass Hamiltonian, here we will choose to work with the BenDaniel and Duke proposal. Physical arguments supporting this choice were put forward by L\'evy-Leblond \cite{32}.

We now consider the complete quantum Hamiltonian of a particle with position-dependent mass
\begin{equation}\label{8}
  \hat{H}=\hat{p}\frac{1}{2m\left(x\right) }\hat{p}+V\left(x \right).
\end{equation}
We then use the transformation
\begin{equation}\label{9}
	 	\hat{H}_\mathrm{T}=\hat{T}^\dagger \hat{H} \hat{T},
\end{equation}
with
\begin{equation}\label{10}
    \hat{T}=\exp\left\lbrace -\frac{i}{2}\left[ \hat{p} g\left(\hat{x} \right)  +  g\left(\hat{x} \right) \hat{p} \right] \right\rbrace,
\end{equation}
where $g(x)$ is a well behaved function that will depend on position. Using the Hadamard lemma \cite{30}, one can show that the momentum operator transforms according to
\begin{equation}\label{11}
    \hat{T}^\dagger \hat{p} \hat{T} = \frac{1}{2}\left[\hat{p}G(x)+G(x)\hat{p} \right],
\end{equation}
where
\begin{equation}\label{12}
     G(x)=\sum_{k=0}^{\infty}\frac{(-1)^kG_k}{k!},
\end{equation}
for which
\begin{equation}\label{13}
      G_{k+1}(x)=g^2(x)\frac{d}{dx}\frac{G_k(x)}{g(x)},\qquad G_0=1.
\end{equation}
On the other hand, for the position operator, we obtain
\begin{equation}\label{14}
	 	\hat{T}\hat{x}\hat{T}^\dagger=x+F(x), \qquad  \hat{T}^\dagger\hat{x}\hat{T}=x+f(x),
\end{equation}
where
\begin{equation}\label{15}
	 	F(x)=\sum_{k=1}^{\infty}\frac{(-1)^kf_k(x)}{k!}, \qquad f(x)=\sum_{k=1}^{\infty}\frac{f_k(x)}{k!},
\end{equation}
with
\begin{equation}\label{16}
	 	f_1(x)=g(x), \qquad  f_{k+1}(x)=g(x)\frac{df_k(x)}{dx}.
\end{equation}
From equation (\ref{11}), we note that
\begin{equation}\label{17}
	 	\hat{T}\hat{p}^2\hat{T}^\dagger=\hat{p}G\hat{p}-\frac{1}{4}\frac{d^2G^2}{dx^2}+\left( \frac{dG}{dx}\right)^2 .
\end{equation}
From the above equations, we can then write
\begin{equation}\label{18}
	 	\hat{H}_\mathrm{T}=\hat{T}^\dagger\hat{H}\hat{T}=\frac{\hat{p}^2}{2}+W(x),
\end{equation}
where the transformed potential $W(x)$ is given by
\begin{equation}\label{19}
	 	W(x)=\tilde{V}\left[x+f(x) \right],
\end{equation}
and where
\begin{equation}\label{19b}
  \tilde{V}(x)=V(x)+\frac{1}{8}\frac{d^2 G^2}{d x^2}-\frac{1}{8} \left(\frac{dG}{dx}\right)^2.
\end{equation}
Up to this point, we have succeeded in eliminating the position dependency of the mass. Note that both Hamiltonians, $\hat{H}$   and $\hat{H}_\mathrm{T}$ have the same sets of eigenvalues since they are related by a similarity transformation. Therefore, by finding the eigenvalues of $\hat{H}_\mathrm{T}$  we can directly obtain the eigenvalues corresponding to the position dependent mass Hamiltonian $\hat{H}$. 

\section{An example}
Let us consider a mass that decays with the position in an exponential-like fashion; i.e., let
\begin{equation}\label{20}
	 	m(x)=\frac{1}{\left( 1+\alpha \beta e^{\beta x} \right)^2}.
\end{equation}
Figure 1, depicts this mass dependence on position when $\alpha=1$ , and for three different values of the parameter $\beta$.
\begin{figure}[h!]
   \centering
   \includegraphics[width=0.75\linewidth]{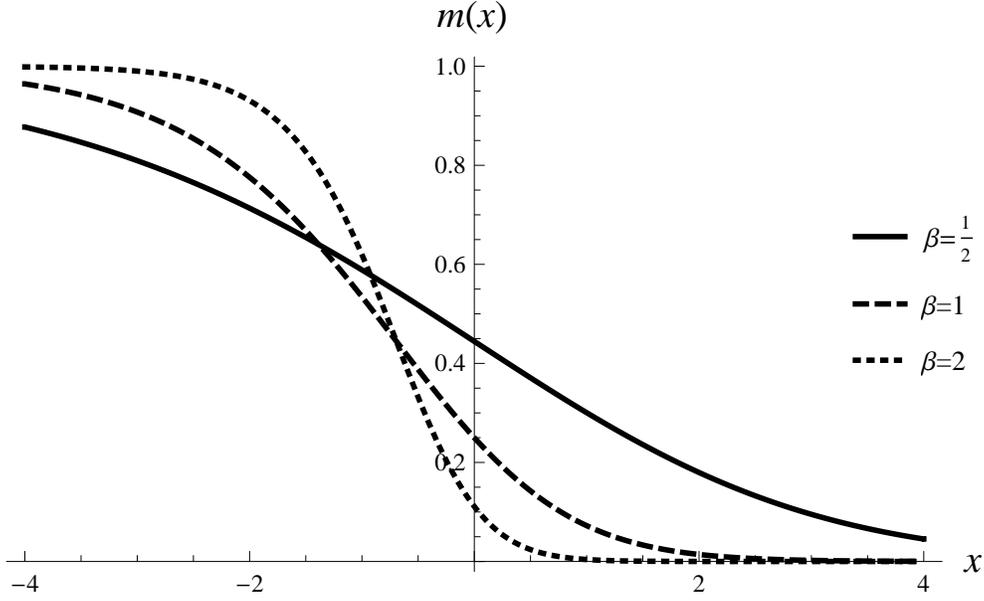}
   \caption{Plot of the mass function (\ref{20}) for $\alpha=1$ , and $\beta=\dfrac{1}{2}, 1, 2$.}
\end{figure} 

This particular dependence of the mass on position suggests the auxiliary function $g(x)=\alpha e^{\beta x} $, in which case the similarity transformation takes the form
\begin{equation}\label{21}
	 	\hat{T}=\exp\left[ -i\frac{\alpha}{2}\left( \hat{p}e^{\beta x}+e^{\beta x}\hat{p}\right) \right] .
\end{equation}
From here one finds that
\begin{equation}\label{22}
	 	f\left( x \right)=-\frac{1}{\beta}\ln\left(1-\alpha \beta e^{\beta x} \right),  
\end{equation}
and
\begin{equation}\label{23}
	 	G\left(x \right)=1+\alpha \beta e^{\beta x},
\end{equation}
that is consistent with $G^2=\dfrac{1}{m}$. \\
With this particular choice for  a position dependent mass (\ref{20}), we also choose the following potential
\begin{equation}\label{24}
	 	V\left(x \right)=a_0+a_1 e^{-\beta x}+a_2 e^{-2 \beta x}+a_3 e^{\beta x}+a_4 e^{2 \beta x},
\end{equation}
with real arbitrary coefficients. If $a_2=\dfrac{a_1^2}{4a_0}$, $a_3=\frac{1}{4}\alpha \beta^3$, and $a_4=-\frac{3}{8}\alpha^2\beta^4$, the transformed potential function $W(x)$ is given by the Morse potential
\begin{equation}\label{25}
	 	W\left( x \right) = D_\mathrm{e}\left[ 1-e^{-\beta\left(x-\gamma \right)}\right]^2,
\end{equation}
where $D_\mathrm{e}=\dfrac{\left( 2 a_0-\alpha\beta a_1\right) ^2}{4 a_0}$ and $\gamma=\dfrac{1}{\beta}\ln\left( \dfrac{a_1}{\alpha \beta a_1-2a_0}\right)$.\\

In Figure 2, we plot the original potential (\ref{24}) as a function of the position for the same values as in Figure 1, when $a_0=a_1=1$  .
\begin{figure}[h!]
  \centering
  \includegraphics[width=0.75\linewidth]{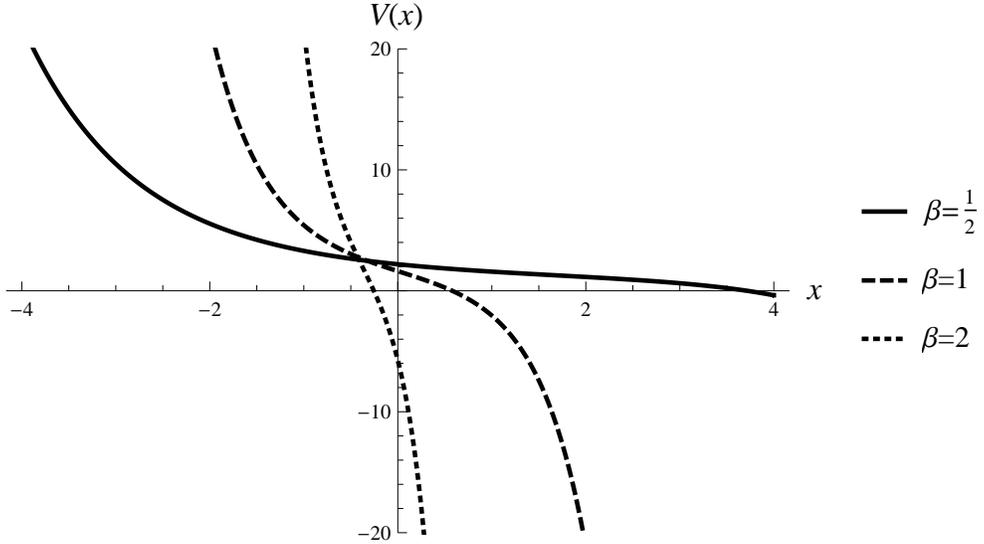}
  \caption{Plot of the potential (\ref{24}) for $\alpha=1$ , $\beta=\dfrac{1}{2}, 1, 2$, and $a_0=a_1=1$.}
\end{figure}
The solution of the transformed equation, can now be obtained given that the mass is constant and the potential involved is of the Morse type that is known to admit analytical solutions.

\section{Conclusions}
By means of a squeeze-like unitary transformation, we have related a position dependent mass Hamiltonian to a Hamiltonian with constant mass. By doing so, we can use standard methods for analyzing quantum mechanical problems of this type. Importantly, the eigenvalues of the transformed Hamiltonian are the same as those associated with the original position dependent problem. Meanwhile the eigenfunctions of these two Hamiltonians are related by a similarity transformation-given by the squeeze-like operator (\ref{10}).

\end{document}